\documentclass{aa403}

\input epsf 

\begin{document}

\sloppy

\thesaurus{3 (9.03.2,13.07.2)}

\title{A search for TeV gamma-ray emission from SNRs, pulsars and 
unidentified GeV sources in the Galactic plane in the
longitude range between $-2^\circ$ and $85^\circ$}
\titlerunning{Point sources in the Galactic plane}
\authorrunning{F. Aharonian et al.}

\author{F.A.~Aharonian\inst{1},
A.G.~Akhperjanian\inst{7},
M. Beilicke\inst{4},
K.~Bernl\"ohr\inst{1},
H.~Bojahr\inst{6},
O.~Bolz\inst{1},
H.~B\"orst\inst{5},
T.~Coarasa\inst{2},
J.L.~Contreras\inst{3},
J.~Cortina\inst{2},
S.~Denninghoff\inst{2},
V.~Fonseca\inst{3},
M.~Girma\inst{1},
N.~G\"otting\inst{4},
G.~Heinzelmann\inst{4},
G.~Hermann\inst{1},
A.~Heusler\inst{1},
W.~Hofmann\inst{1},
D.~Horns\inst{1},
I.~Jung\inst{1},
R.~Kankanyan\inst{1,7},
M.~Kestel\inst{2},
J.~Kettler\inst{1},
A.~Kohnle\inst{1},
A.~Konopelko\inst{1},
H.~Kornmeyer\inst{2},
D.~Kranich\inst{2},
H.~Krawczynski\inst{1,}$^\%$,
H.~Lampeitl\inst{1},
M. Lopez\inst{3},
E.~Lorenz\inst{2},
F.~Lucarelli\inst{3},
O.~Mang\inst{5},
H.~Meyer\inst{6},
R.~Mirzoyan\inst{2},
A.~Moralejo\inst{3},
E.~Ona\inst{3},
M.~Panter\inst{1},
A.~Plyasheshnikov\inst{1,}$^\S$,
G.~P\"uhlhofer\inst{1},
G.~Rauterberg\inst{5},
R.~Reyes\inst{2},
W.~Rhode\inst{6},
J.~Ripken\inst{4}
A.~R\"ohring\inst{4},
G.P.~Rowell\inst{1},
V.~Sahakian\inst{7},
M.~Samorski\inst{5},
M.~Schilling\inst{5},
M.~Siems\inst{5},
D.~Sobzynska\inst{2,}$^*$,
W.~Stamm\inst{5},
M.~Tluczykont\inst{4},
H.J.~V\"olk\inst{1},
C.A.~Wiedner\inst{1},
W.~Wittek\inst{2}}

\institute{Max Planck Institut f\"ur Kernphysik,
Postfach 103980, D-69029 Heidelberg, Germany \and
Max Planck Institut f\"ur Physik, F\"ohringer Ring
6, D-80805 M\"unchen, Germany \and
Universidad Complutense, Facultad de Ciencias
F\'{i}sicas, Ciudad Universitaria, E-28040 Madrid, Spain
\and
Universit\"at Hamburg, Institut f\"ur
Experimentalphysik, Luruper Chaussee 149,
D-22761 Hamburg, Germany \and
Universit\"at Kiel, Institut f\"ur Experimentelle und Angewandte Physik,
Leibnizstra{\ss}e 15-19, D-24118 Kiel, Germany\and
Universit\"at Wuppertal, Fachbereich Physik,
Gau{\ss}str.20, D-42097 Wuppertal, Germany \and
Yerevan Physics Institute, Alikhanian Br. 2, 375036 Yerevan,
Armenia\\
\hspace*{-4.04mm} $^\S\,$ On leave from
Altai State University, Dimitrov Street 66, 656099 Barnaul, Russia\\
\hspace*{-4.04mm} $^\%$ Now at Yale University, P.O. Box 208101, New Haven, CT 06520-8101, USA\\
\hspace*{-4.04mm} $^*$ Home institute: University Lodz, Poland\\
}
\mail{Hubert Lampeitl, \\Tel.: (Germany) +6221 516 528,\\
email address: lampeitl@daniel.mpi-hd.mpg.de}
\offprints{Hubert Lampeitl}

\date{Received 6 March 2002; accepted 16 August 2002}

\maketitle

\begin{abstract}

Using the HEGRA system of imaging atmospheric Cherenkov telescopes,
one quarter of the Galactic plane
($-2^\circ < l < 85^\circ$) was 
surveyed for TeV gamma-ray emission from point sources and moderately
extended sources (\O$\;\le0.8^\circ$).
The region covered includes 86 known pulsars (PSR), 63 known 
supernova remnants (SNR) and nine GeV sources,
representing a significant fraction of the known populations. 
No evidence for emission of TeV gamma radiation was detected, and
upper limits range from 0.15 Crab units up to several Crab
units, depending on the observation time and zenith angles covered.
The ensemble sums over selected SNR and pulsar subsamples and over the
GeV-sources yield 
no indication for emission from these potential sources. 
The upper limit for the SNR population is at the level of 6.7\% of the
Crab flux and for the pulsar ensemble at the level of 3.6\% of the Crab flux.

\keywords{Gamma rays: observations, ISM: cosmic rays, pulsars:
  general, supernova remnants}

\end{abstract}

\section{Introduction}
\begin{figure}
\begin{center}
\mbox{
\epsfxsize7.8cm
\epsffile{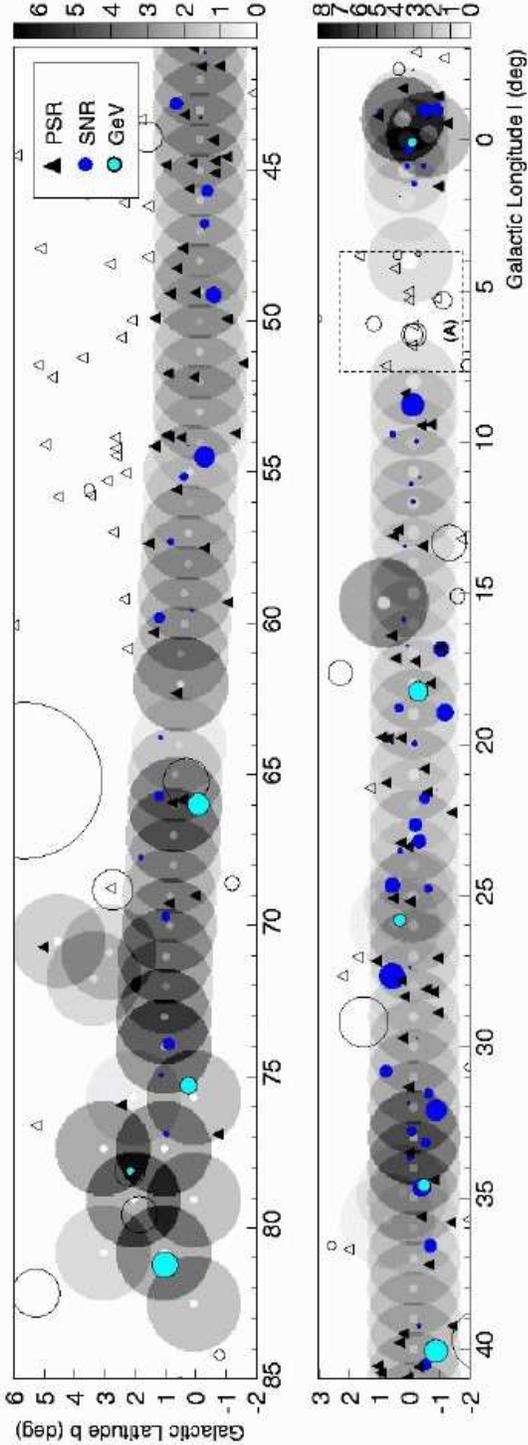}}
\caption{Observation time in hours used for the individual scan points. 
The large gray circles indicate the individual pointings and correspond
the used FoV of the telescope system.
Positions of potential TeV gamma-ray sources are marked by symbols;
a filled symbol indicates a potential source for which we
give an upper limit. The size of the circles for SNRs and GeV sources
corresponds to the size of the source. Objects in the dashed box labeled (A)
are excluded from further analysis (for explanation see text).
}
\label{obs_time}
\end{center}
\end{figure}
\begin{figure}
\begin{center}
\mbox{
\epsfxsize8.5cm
\epsffile{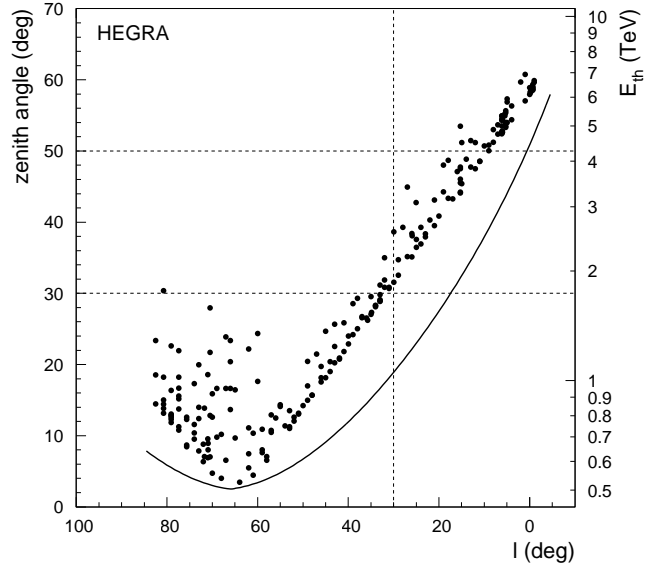}}
\caption{Correlation between Galactic longitude and zenith angle under
which the individual scan points were observed. Each dot represents 
a data taking period of 20 or 30 min. The solid line indicates the 
approximated energy threshold in TeV (right axis) as a function of 
Galactic longitude (Konopelko et al., 1999b).}
\label{fig_alt_l_correlation}
\end{center}
\end{figure}
Systems of imaging atmospheric Cherenkov telescopes such as the 
HEGRA stereoscopic telescope system (Daum et al. 1997, Konopelko et al. 1999a),
allow to reconstruct 
the directions of air showers over the full field of view,
with a radius of about 2$^\circ$ in the case of HEGRA, and can 
therefore be used for
sky surveys (P\"uhlhofer et al. 1999, Aharonian et al. 2001b). 
Here, we report on a survey of one quarter of the Galactic disc
ranging from the Galactic center (l$\;\approx 0^\circ$) to the Cygnus region
(l$\;\approx 83^\circ$). 
The latitude range covered corresponds in most parts of the survey
to the FoV of the HEGRA telescope system and ranges from $-1.7^\circ$ to
$1.7^\circ$ (for more details see Fig. \ref{obs_time}).
The motivation for this survey was to search for
gamma-ray point sources and moderately extended sources in the TeV energy range. 
Most of the 
potential Galactic gamma-ray
sources like supernova remnants (SNR) (Green, 1998) and pulsars (PSR)
(Taylor, 1993) are the
remnants of young massive (Population I) stars and thus
cluster along the Galactic plane and concentrate towards
the Galactic center.
This picture is supported by earlier $\gamma$-ray surveys carried 
out with the COS B
satellite (Swanenburg et al., 1981) and with the EGRET
instrument (Hartmann et al. 1999, Lamb and Macomb, 1997) in the GeV 
range revealing an enhancement of $\gamma$-ray sources along the
Galactic plane.
Both types of objects - SNRs and pulsars -
are almost certainly particle accelerators and 
emitters of high-energy gamma radiation. Theoretical
models predict typical gamma-ray fluxes from the majority
of these objects are below the detection thresholds of the
current generation of Cherenkov instruments (see, e.g.
Drury et al. 1994, Aharonian et al. 1997 and 
Berezhko \& V\"olk 2000a). Until now only three 
SNRs - SN1006 (Tanimori et al. 1998), RX J1713.7-3946 
(Muraishi et al. 2000) in the southern hemisphere and Cas-A
(Aharonian et al. 2001a) in the northern hemisphere show evidence
for TeV gamma-ray emission. For SN1006 a flux at the level of 70\%
of the Crab flux\footnote{To keep calculations simple, we give fluxes
in units of the Crab Nebula flux (so called CU). For the Crab we take a value of:\\
$F(>E) = 1.75\times10^{-11} \left(\frac{E}{1~TeV}\right)^{-1.59}
\mathrm{ph~cm^{-2} s^{-1}}$\\ (Aharonian et al. 2000) }is reported, 
for RX J1713.7-3946 at the level of 80\%
and for Cas-A at the level of 3.3\%. 
For the individual shell type SNRs, $\gamma$-Cygni, IC-433, W44, W51
upper limits at the level of 20\% to 30\% of the Crab flux are given in 
Buckley (1998) and V\"olk (1997).
For the SNR W28 an upper limit of 70\% of the Crab flux is given in
Rowell et al. (2000) and for
Tycho an upper limit of 3.3\% of the Crab is given in 
Aharonian et al. (2001c).
Three pulsars - the Crab Nebula (Weekes 1989), PSR1706-44 (Kifune et
al. 1995) at the level of 60\% and Vela (Yoshikoshi et al. 1997)
at the level of 70\% of the Crab flux have been reported as 
gamma-ray emitters in the TeV regime. 
For a review of observations 
and theoretical predictions on Galactic gamma-ray sources see, 
e.g, Aharonian, 1999c.
In addition to pulsars and SNRs, many unidentified
GeV sources (Lamb \& Macomb 1997) lie in the Galactic plane.\\
Both the lack of knowledge of the 
individual source parameters as well as the
approximations used in the modeling result in large
uncertainties in the predictions for individual objects by an order
of magnitude or more. Hence it is desirable to observe a larger sample
of source candidates beyond the few most promising representatives of
each class. Given the
density of source objects, a survey of the inner part of the
Galactic plane provides an efficient
way to search for gamma-ray emission and to average over the
potential source populations. \\
With the HEGRA telescope system such a survey was conducted. The 
range of the survey, $-2^\circ < l < 85^\circ$, was chosen either by
visibility conditions and by the density of potential gamma-ray emitters.
From the location of the HEGRA telescope system
at $28^\circ 45'$ N, observation conditions are best for Galactic 
longitudes around $65^\circ$. The Galactic center can only be observed
at large zenith angles around $60^\circ$, and most parts of the
Galactic plane with negative longitudes are virtually inaccessible.
\section{The HEGRA IACT system}
The HEGRA stereoscopic system (Daum et al. 1997, Konopelko et al. 1999a)
of imaging atmospheric Cherenkov telescopes (IACTs)
is located on the Canary Island of La Palma, 
on the site of the Observatorio del Roque de los Muchachos,
at $28^\circ 45'$ N, $17^\circ 53'$ W, 2200 m a.s.l. 
The stereoscopic telescope system
consists of five telescopes (CT2-CT6). One additional telescope
(CT1, Mirzoyan et al. 1994) is operated in stand-alone mode.
The system telescopes are arranged on the corners and in the center of a
square with 100 m side length.
Each is equipped with a tessalated 8.5~m$^2$ mirror
of 5 m focal length, and a camera with 271 photomultiplier
pixels in the focal plane. The field of view of each camera
has a diameter of 4.3$^\circ$ with pixel diameters corresponding 
to 0.25$^\circ$. 
The telescope system is triggered when in at least two
cameras two neighboring pixels show a signal above $\approx 7$
photoelectrons. Signals from the cameras are recorded using 
a 120 MHz Flash-ADC system, which is read out after a system trigger.
Details of the camera hardware and of the trigger system
are given in Hermann (1995) and Bulian et al. (1998). 
The pointing uncertainty of the telescope system is below
1 arcmin (P\"uhlhofer et al. 1997). 
On the basis of
the stereoscopic analysis of Cherenkov images, shower
directions can be reconstructed with an accuracy of 0.1$^\circ$, 
and shower energies with a resolution of 20\% or better
(Daum et al. 1997; Aharonian et al. 1999a,b).
The energy threshold is 500 GeV for vertical incidence of gamma-rays, 
and increases to 0.9 TeV at 30$^\circ$, to 1.8 TeV at 45$^\circ$ 
and to 5 TeV at 60$^\circ$ (Konopelko et al. 1999b). 
Cosmic ray showers are suppressed exploiting
the width of Cherenkov images. A ``mean scaled width''
$\bar{w}$ is defined by 
scaling the observed widths to the expected widths for gamma-ray 
images, which depend on the intensity of the images,
the distance to the shower core and the zenith angle, and averaging over
telescopes. Gamma-rays cause a peak in $\bar{w}$
at 1, with a Gaussian width of about 0.l.
Nucleonic showers have larger $\bar{w}$ values, peaking around 1.7.
While more sophisticated identification schemes 
(e.g., Daum et al. 1997, Lampeitl \& Konopelko 1999, Sch\"afer et al. 2001) 
can reach slightly better sensitivity, the default (and most stable)
analysis schemes are based on cuts in $\bar{w}$ ($\bar{w} <
1.1...1.3$), combined with an angular cut relative to the source of 
about 0.15$^\circ$ in case of a point source. 
\section{The dataset}
Data used in this survey were taken from June to September in
1997 and from June to August in 1998 with a 4-telescope system
\footnote{CT2 was incorporated into the HEGRA-Telescope-System in May 1999.}.
In total 176~h of observation time distributed over 92 separated
locations along the Galactic plane were obtained. 
The general layout of the survey is illustrated in Fig. \ref{obs_time}.
The observations
mainly followed the galactic equator with a spacing between individual
scan positions of $1^\circ$, allowing an overlap of the FoV between
different scan points. In the outer region of the Galactic plane ($l>54^\circ$)
the survey points are slightly displaced to the north following  
the density of interstellar matter. In the Cygnus region ($l\approx 80^\circ$)
the survey points 
cover a larger range in Galactic latitude $b$, because the
distribution of matter is much broader in this region, at the expense
of a reduced overlap between adjacent points.
The observation schedule was optimized
such that individual survey points were observed near culmination,
resulting in the smallest zenith angle accessible. This observation 
strategy leads to a
strong correlation between zenith angle and Galactic longitude. As
shown in Fig. \ref{fig_alt_l_correlation} this correlation also
implies a variation of the effective energy threshold with 
Galactic longitude $l$. \\
A quality selection of the data sets was based primarily
on the average trigger rate of the telescope system. Data 
affected by bad weather conditions and technical problems
were excluded from further analysis.
The remaining data set encompasses 115~h of observation time.\\
The analysis of Cherenkov images could potentially suffer from
variations in the sky brightness over the scan region. Since 
the readout electronics of the telescopes is AC coupled,
a star illuminating a pixel will not cause baseline shifts,
but it will still result in increased noise in that pixel.
The observed region contains only one star brighter than 3.5 mag at
l=78.15$^\circ$,b=1.87$^\circ$ with a magnitude $m_v = 2.23$ mag. Since the 
galactic equator is obscured by interstellar dust, in most regions
of the scan the Galactic background light is negligible. 
Only the region around 
l=6$^\circ$ (marked in Fig. \ref{obs_time} with A) shows strong influence 
of background light caused by the star clusters M8 and NGC6530.
Data taken at this location show a strong inhomogeneity over the FoV
and were excluded from this analysis. \\
\begin{figure}
\begin{center}
\mbox{
\epsfxsize9.0cm
\epsffile{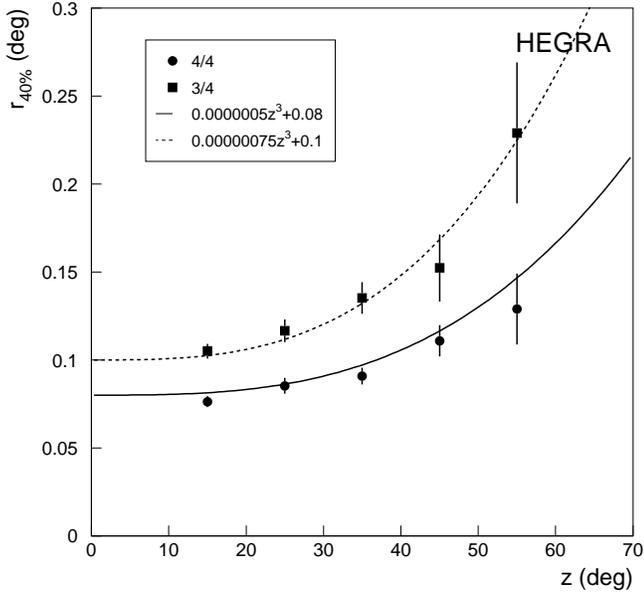}}
\caption{Angular resolution ($r_{40\%}$) derived from the Crab data 
as a function of
zenith angle $z$, after a cut on mean scaled width of $\bar{w}<1.1$. 
Shown is the angular resolution for events with four telescopes out
of four $(4/4)$ used in the analysis and three out of four $(3/4)$.
The lines indicate the parametrisation used for further analysis.}
\label{fig_angular_res}
\end{center}
\end{figure}
\begin{figure}
\begin{center}
\mbox{
\epsfxsize9.0cm
\epsffile{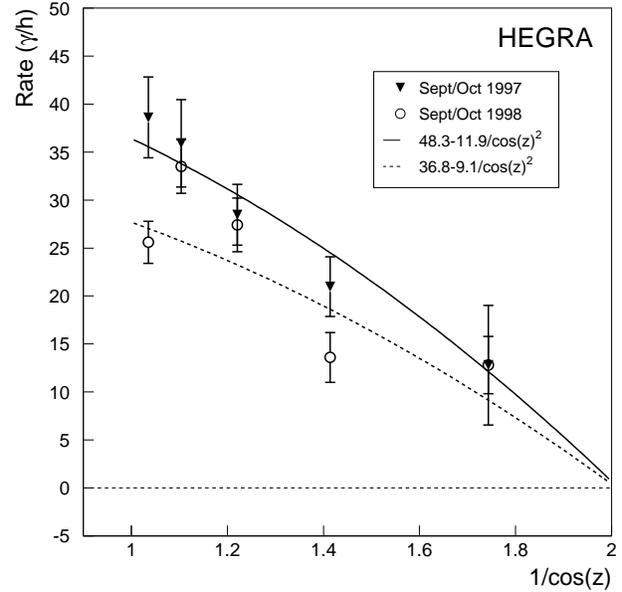}}
\caption{Observed photon rate for the Crab Nebula in 1997 and 1998 as a
function of $1/\cos(z)$ for 3- and 4-telescope events. An angular 
cut optimized for the angular resolution of the telescope system was
applied. The lines indicate the parametrisation used in the analysis.}
\label{fig_crab_rate}
\end{center}
\end{figure}
\begin{figure}[t]
\begin{center}
\mbox{
\epsfxsize8.5cm
\epsffile{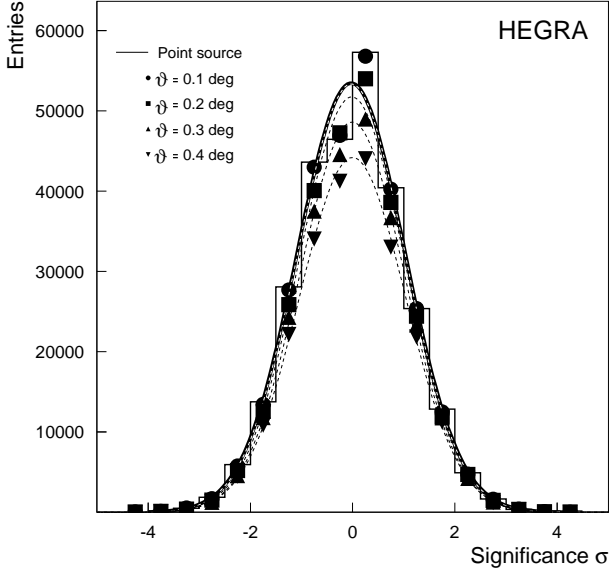}}
\caption{Distribution of significances for the grid points and for
different search radii.
No point above 4.5 $\sigma$ is detected. 
Taking the amount of trials into account, this result is fully
compatible with background noise. The curves shown
indicate Gaussian fits to the data points. The curves for point sources
and for sources with an extension of $0.1^\circ$ are nearly identical.
}
\label{fig_sig_dis}
\end{center}
\end{figure}
For reference and comparison, observations of the Crab Nebula
in September and October 1997 and September and October 1998 
were used, and were subject to identical selection criteria. 
In total 114 h of Crab data covering the zenith angle range 
from $10^\circ$ to $60^\circ$ were used. The angular resolution
of the telescope system was derived by fitting a two dimensional
Gaussian distribution to the spatial distribution of the gamma ray events
from the Crab Nebula. As a measure for the angular resolution
the $40\%$ containment radius $r_{40\%}$ of the two dimensional 
distribution was taken equivalent to the Gaussian width of the
projected one dimensional angular distribution. The resulting angular
resolution for different zenith angles is shown in 
Fig. \ref{fig_angular_res}. For weak sources the optimum significance 
$S/\sqrt{B}$ is obtained by cutting the distribution at a
radius of $1.6 \cdot r_{40\%}$. After applying the angular cut
the detection rate for the Crab Nebula was determined.
The result is shown in Fig. \ref{fig_crab_rate}. Since aging
processes of the PM tubes and different HV settings in 1997 and 1998
affected the detection rate for gamma-rays, 
individual calibration curves were derived for these years 
\footnote{We note that the different $\gamma$-ray detection rates for different 
years indicate an change in the energy threshold of order of 15\%.}.
The off axis sensitivity of the telescope system was determined by 
Monte Carlo simulations (for more details see Aharonian et al. 2001b).
Simulations carried out for zenith angles of $20^\circ$ and $30^\circ$
show that the detection rate for off axis gamma rays could be well
described for $\vartheta < 1.7^{\circ}$ by a dependence of the 
form $(1-0.1\vartheta^2)$ where $\vartheta$ is the inclination of the 
shower axis with respect to the telescope pointing in degrees. The
simulation show that the detection 
rate is reduced by 10\% for an inclination of $1^\circ$ and by $30\%$ at 
$1.7^\circ$. For larger inclinations border effects by truncated images
start to deteriorate the smooth behavior.
\section{Search for gamma-ray sources}
The cuts on the telescope images and the shower reconstruction
follow earlier work (see e.g. Aharonian et al. 1999a). 
In particular, only images with at least
40 photoelectrons were accepted and the centroid of the 
image had to be within  1.7$^\circ$ from the camera center, in
order to exclude truncated images. Showers with reconstructed
cores up to 300 m from the center telescope were accepted.
Since angular resolution and gamma/hadron-separation improves
with the number of telescopes used for shower reconstruction,
only events with three or more triggered telescopes were included 
in the analysis.\\
The search for gamma-ray sources was carried out on a grid  
of 0.03125$^\circ$ spacing, well below the angular resolution 
of the telescope system. For each grid point 
the number of events located within a search radius $r_s$ were counted. 
The background was determined by three control regions rotated 
by 90$^\circ$, 180$^\circ$ and
270$^\circ$ with respect to the pointing of the telescopes.
Each potential source location was analyzed assuming a point source,
as well as extended sources of radii $r = 0.1^\circ, 0.2^\circ,
0.3^\circ$ and $0.4^\circ$ and taking in addition into account the zenith 
angle dependent angular resolution of the system (see Fig. \ref{fig_angular_res}). 
By applying this scheme,
the number of background events could not be evaluated for sources on
or near the center of the FoV, and hence the inner part of the FoV of 
radius $r_s$ was excluded from this
analysis. This results in a minor loss of sky coverage, since neighboring
scan points cover in most part of the scan the excluded regions. 
The radius of the FoV for sources was limited to $1.7^\circ-r_s$ to exclude 
the influence of the camera borders.
With the event counts in the search region and the three control regions
the significances $\sigma$ were calculated according
to Li \& Ma (1993) and upper limits according to
O. Helene (1994) for each grid point. To derive an upper limit on the 
flux the result was divided by the expected number of events for a
Crab-like source observed for the same time
at the same zenith angle and at the same inclination with respect to the
pointing of the telescopes. \\
 The distribution of significances for different search radii
is shown in Fig. \ref{fig_sig_dis}. All distributions are well fitted
by a Gaussian distribution of mean value 0 and variance of 1,
indicating that the background estimation is reliable.
In none of the distributions a
point above 4.5$\sigma$ is evident. Taking the number of trials into
account the result is fully compatible with background noise.\\
In the following we give limits for various individual sources in
the range of the survey.
For the nine GeV sources located in the scan region results are given in 
Table \ref{tab_gev}. The source GeV J1746-2854 coincides with the 
Galactic center.
The limit of 8.7 Crab can thus be interpreted as an upper limit on the
emission from the Galactic center. We note that the Galactic center
is observed around a zenith angle of $60^\circ$ and that at such large
zenith angles the detection rate for gamma-rays is quite low (see
Fig. \ref{fig_crab_rate}) and angular resolution is degraded by more 
than a factor of 2 (see Fig. \ref{fig_angular_res}). 
Systematic effects might well dominate over statistics and a systematic error
on a level of 50\% on this limit can not be excluded.\\
Results for 19 SNRs out of 63 with an estimated distance of less
than 10 kpc, a diameter smaller than $0.8^\circ$
and observed under zenith angles smaller than $45^\circ$
are given in
Table \ref{tab_snr_selected}.
Results for 18 pulsars selected from 86
with a characteristic age less than $10^6$ years, a distance smaller than 10 kpc 
and which were observed under zenith angles smaller than $45^\circ$ are given in 
Table \ref{tab_psr_selected}. Results on the remaining SNRs and pulsars
are given in Table \ref{tab_snr2}.
Limits obtained range from 7\% of the Crab flux up to
several Crab units, depending on zenith angle, accumulated observation 
time and search radius.
\section{Ensemble limits}
Even if individual sources show no indication for TeV gamma-ray
emission,
one can try to increase the experimental sensitivity by
considering whole ensembles of sources ("source stacking").
For given source classes, the source-region and background-region
counts were accumulated and a significance and a limit was derived for
each ensemble.
\subsection{Unidentified GeV-sources}
\begin{table*}[t]
\begin{center}
\caption{Results for GeV sources taken from the catalogue of Lamb \&
Macomb (1997). $\sigma_{pos}$ indicates the 95\% error box on the
source location, $r$ the assumed radius of the source (P = point like
source). 
T denotes the
observation time, ON the number of events in the source bin and OFF
the number of events in the control regions. $\sigma$ gives the
significance calculated according to Li \& Ma (1983) with
$\alpha=1/3$. 
Exp. gives the number of expected events for a Crab like source
observed at the same
zenith angle the same observation time and at the same inclination with
respect to the telescope pointing. E$_{th}$ gives the approximate energy threshold
in TeV. $F^{99\%}$ is the derived upper
limit in units of the Crab flux (CU). Possible coincidences of GeV-sources
with SNRs are indicated in the last column, as given by 
Romero et al. (1999). The last line gives the
sum over the GeV sources 3 to 9 which were observed under zenith
angles $z<45^{\circ}$.  
}
\vspace{0.3cm}
\scriptsize
\begin{tabular}{|r|l|r|r||r|r|r|r|r|r|r||l|}
\hline
 & Name & $\sigma_{pos}$ & $r$ &      T & ON & OFF & $\sigma$ & Exp.  & E$_{th}$ & $F^{99\%} $ & Remark \\
 &      & [$^{\circ}$] & [$^{\circ}$]& [h]&    &     &          &       &  [TeV] & [CU]          & \\
\hline
\hline
1 & GeV J1746-2854 & 0.16 & P  & 7.1 & 1183 & 3443 & 0.9 & 15.0 & 4.5 & 8.67 &Gal. cent., G0.0+0.0, G0.5+0.0\\
2 &GeV J1825-1310 & 0.32 & 0.4& 1.0 & 79   & 201  &1.23 & 20.4  & 1.7 & 1.87 & \\
\hline
3 & GeV J1837-0610 & 0.20 & 0.2& 3.6 & 51   & 170  &-0.67&  86.2 & 1.3 & 0.24 & \\
4 & GeV J1856+0115 & 0.21 & 0.3& 3.2 & 49   & 216  &-2.53& 101.1 & 0.9 & 0.14 & G34.7-0.4 (W44)\\
5 & GeV J1907+0557 & 0.36 & 0.4& 2.1 & 65   & 169  &0.97 &  67.7 & 0.8 & 0.48 &
G39.2-0.3 \\
6 & GeV J1957+2859 & 0.36 & 0.4& 3.0 & 32   & 150  &-2.40&  78.4 & 0.6 & 0.15 & \\
7 & GeV J2020+3658 & 0.28 & 0.3& 1.6 & 27   & 65   &0.95 &  56.1 & 0.6 & 0.38 & \\
8 & GeV J2020+4023 & 0.14 & 0.2& 6.3 & 36   & 102  &0.29 & 181.9 & 0.6 & 0.12 & G74.9+1.2\\
9 & GeV J2035+4214 & 0.42 & 0.4& 2.1 & 59   & 172  &0.19 &  74.2 & 0.6 & 0.35 & \\
\hline
& $\Sigma_{3-9}$  &      &    & 21.9 & 319  &1044 &  -1.4  & 645.6& - & 0.057&  \\
\hline
\end{tabular}
\label{tab_gev}
\end{center}
\end{table*}
\begin{table*}[t]
\begin{center}
\caption{List of selected SNR from Green (1998). Type indicates the
  morphology according to Green (C = Composite, S = shell type, ? =
  some uncertainties). 
$r_{SNR}$ denotes the radius of the SNR in degrees. $r$ indicates the
assumed source size as taken in the analysis (P = point like source). 
$d$ is the distance to the individual remnant in kpc as derived from the 
radio-surface-brightness-to-diameter relationship ($\Sigma$-$D$) as given
in Case \& Bhattacharya (1998). 
$T/d^2$ is the weight of the individual remnant as used in 
Eq. \ref{equ_ensemble_sum}.
The other columns are labeled as in Table \ref{tab_gev}. The last 
row gives the sum over the ensemble of SNRs. 
$F_{DAV}$ is the expected hadronic flux from the ensemble
calculated according to Drury et al. (1994). For further explanations
see text.}
\vspace{0.3cm}
\scriptsize
\begin{tabular}{|r|l|c|r|r|r||r|r|r|r|r|r|r|r|}
\hline
Nr & Name & Type & $r_{SNR}$ & $r$ & $d$ & T & ON & OFF & $\sigma$ & Exp. & $T/d^2$ & E$_{th}$ & $F^{99\%}$ \\
   &      &      & $[^o]$  & $[^o]$  &[kpc]& [h]&   &     &          & 
     & [h/kpc$^2$] & [TeV] & [CU]   \\
\hline
\hline
1  &G021.5-00.9 & C  & 0.01 & P   & 6.3 & 1.6 & 23 &  44 & 1.7  & 33.0 & 0.040 & 1.5 & 0.70 \\
2  &G021.8-00.6 & S  & 0.17 & 0.2 & 2.9 & 2.6 & 34 &  85 & 0.9  & 59.6 & 0.309 & 1.5 & 0.39 \\
3  &G022.7-00.2 & S? & 0.22 & 0.3 & 3.7 & 1.2 & 32 &  69 & 1.5  & 29.2 & 0.088 & 1.4 & 0.89 \\
4  &G023.3-00.3 & S  & 0.22 & 0.3 & 2.7 & 1.2 & 29 &  76 & 0.6  & 29.9 & 0.165 & 1.4 & 0.69 \\
5  &G024.7-00.6 & S? & 0.12 & 0.2 & 7.4 & 3.3 & 41 & 144 & -0.9 & 79.9 & 0.060 & 1.3 & 0.23 \\
6  &G030.7+01.0 & S? & 0.20 & 0.3 & 7.9 & 2.1 & 37 & 128 & -0.8 & 56.4 & 0.034 & 1.0 & 0.30 \\
7  &G031.9+00.0 & S  & 0.04 & 0.0 & 7.2 & 4.2 & 21 &  56 & 0.5  &107.2 & 0.081 & 1.0 & 0.16 \\
8  &G032.8-00.1 & S? & 0.14 & 0.2 & 6.3 & 3.1 & 38 &  90 & 1.2  & 84.9 & 0.078 & 1.0 & 0.32 \\
9  &G033.6+00.1 & S  & 0.08 & 0.1 & 7.1 & 5.8 & 38 & 108 & 0.3  &159.9 & 0.115 & 0.9 & 0.14 \\
10 &G034.7-00.4 & C  & 0.29 & 0.3 & 3.3 & 3.7 & 68 & 229 & -0.9 &115.8 & 0.340 & 0.9 & 0.19 \\
11 &G039.2-00.3 & S  & 0.07 & 0.1 & 5.9 & 3.1 & 20 &  62 & -0.1 & 87.5 & 0.089 & 0.8 & 0.18 \\
12 &G040.5-00.5 & S  & 0.18 & 0.2 & 6.1 & 2.1 & 35 &  73 & 1.7  & 65.1 & 0.056 & 0.8 & 0.43 \\
13 &G041.1-00.3 & S  & 0.04 & P   & 6.1 & 3.2 & 16 &  38 & 0.8  & 87.3 & 0.086 & 0.8 & 0.18 \\
14 &G043.3-00.2 & S  & 0.03 & P   & 7.5 & 3.4 & 16 &  50 & -0.1 & 96.3 & 0.060 & 0.7 & 0.14 \\
15 &G045.7-00.4 & S  & 0.18 & 0.2 & 9.1 & 2.6 & 26 &  87 & -0.5 & 80.7 & 0.031 & 0.7 & 0.20 \\
16 &G046.8-00.3 & S  & 0.14 & 0.2 & 6.4 & 2.6 & 23 &  62 & 0.4  & 79.2 & 0.063 & 0.7 & 0.22 \\
17 &G049.2-00.7 & S? & 0.25 & 0.3 & 6.0 & 3.0 & 48 & 129 & 0.6  & 97.3 & 0.083 & 0.6 & 0.27 \\
18 &G054.4-00.3 & S  & 0.33 & 0.4 & 3.3 & 1.5 & 37 & 122 & -0.5 & 47.9 & 0.138 & 0.6 & 0.38 \\
19 &G073.9+00.9 & S? & 0.18 & 0.2 & 6.6 & 2.0 &  5 &  17 & -0.2 & 47.9 & 0.046 & 0.6 & 0.18 \\ 
\hline                                                                       
&$\Sigma_{1-19}$&  &      &  & $<5.2>$&52.3& 587&1669& 1.1  & 1445.0 & 1.962 & - &0.067 \\
\hline                                          
\end{tabular}
\label{tab_snr_selected}
\end{center}
\end{table*}
One can use source stacking under the assumption that all of the
unidentified GeV sources here are of similar type, making any upper
limits meaningful to any source-specific model. 
For seven GeV sources located at a Galactic longitude $l>25^\circ$
the sum is given in Table \ref{tab_gev}. In total 21.9 h of on source 
observation time was accumulated and no indication for 
TeV gamma-ray emission was found.
The significance $\sigma$ is calculated to -1.4 and the upper limit to 5.7\%
of the Crab flux.
\subsection{SNRs}
\begin{table*}[t]
\begin{center}
\caption{List of selected pulsars as taken from Taylor (1993). Pulsars are
  selected by their modeled distance $d_m$ less than 10 kpc, rotation period
  less than 1 s and a characteristic age $\tau = 1/2~p/\dot{p}$ less than $10^6$
  years. $log(\dot{e})$
denotes the logarithm of the spin down luminosity of the pulsar in erg/s. F
gives the spin down energy of the pulsar divided by the square of the
distance $d_m$ in erg s$^{-1}$ kpc$^{-2}$. The last column gives the individual 
contribution of the pulsar 
to the ensemble sum. The ensemble sum is given in the last row.
Other columns are labeled as in Table \ref{tab_gev}. 
Pulsars are treated in the analyses as point sources. The search radius
is chosen to $1.6 \cdot r_{40\%}(z)$ (see Fig. \ref{fig_angular_res}).
For the calculation of the ratio $T_i/T \cdot F_i/F_{Crab}$ the
sum of the total observation time T= 51.0 h is used.}
\vspace{0.3cm}
\scriptsize
\begin{tabular}{|r|l|c|r|r|r|r|r|r||r|r|r|r|r|r|}
\hline
   & Name       & T &  ON &OFF& Exp. & $\sigma$ & $E_{th}$ & F$^{99\%}$ & $\log(\dot{e})$& $d_{m}$ & $\tau$     & $F=\dot{e}/d_m^2$& $\frac{T_i}{T} \cdot \frac{F_i}{F_{Crab}}$ \\
   &            &[h]&     &   &      &          & [TeV]& [CU]       & [erg/s]        & [kpc]   & [$10^5$ y] &                  & \\
      
\hline
\hline
   & Crab       & - &   - &  -&     -&   -    &  -  & -  &38.649 & 2.49 & 0.013 & $7.19\cdot 10^{37}$ &   \\
\hline
 1 & J1830-1059 & 2.1 & 17 & 75 & 45.4  & -1.49 & 1.5 & 0.24 &  34.552 & 3.63 & 1.07 & $2.71\cdot 10^{33}$& $1.54\cdot 10^{-06}$ \\
 2 & J1832-0827 & 2.3 & 31 & 61 & 52.5  &  1.86 & 1.4 & 0.51 &  33.969 & 4.75 & 1.61 & $4.13\cdot 10^{32}$& $2.56\cdot 10^{-07}$ \\
 3 & J1833-0827 & 2.3 & 25 & 71 & 52.9  &  0.23 & 1.4 & 0.34 &  35.766 & 5.67 & 1.47 & $1.81\cdot 10^{34}$& $1.13\cdot 10^{-05}$ \\
 4 & J1835-06   & 3.8 & 22 & 99 & 86.4  & -1.79 & 1.3 & 0.13 &  34.747 & 6.34 & 1.20 & $1.39\cdot 10^{33}$& $1.45\cdot 10^{-06}$ \\
 5 & J1836-1008 & 1.3 & 10 & 15 & 25.0  &  1.64 & 1.4 & 0.61 &  33.416 & 5.40 & 7.58 & $8.94\cdot 10^{31}$& $3.19\cdot 10^{-08}$ \\
 6 & J1841-0425 & 2.6 & 19 & 73 & 57.7  & -0.98 & 1.2 & 0.21 &  34.592 & 5.16 & 4.62 & $1.47\cdot 10^{33}$& $1.04\cdot 10^{-06}$ \\
 7 & J1844-0538 & 2.6 & 17 & 62 & 54.5  & -0.73 & 1.2 & 0.23 &  34.360 & 6.16 & 4.18 & $6.04\cdot 10^{32}$& $4.29\cdot 10^{-07}$ \\
 8 & J1844-0244 & 3.1 & 14 & 58 & 78.7  & -1.12 & 1.1 & 0.14 &  33.703 & 5.99 & 4.81 & $1.41\cdot 10^{32}$& $1.20\cdot 10^{-07}$ \\
 9 & J1845-0434 & 2.6 & 22 & 61 & 55.7  &  0.31 & 1.1 & 0.31 &  34.543 & 4.72 & 6.81 & $1.57\cdot 10^{33}$& $1.11\cdot 10^{-06}$ \\
10 & J1847-0402 & 2.6 & 22 & 32 & 59.1  &  2.53 & 1.1 & 0.43 &  33.980 & 3.13 & 1.83 & $9.75\cdot 10^{32}$& $6.94\cdot 10^{-07}$ \\
11 & J1856+0113 & 3.7 & 19 & 61 & 101.0 & -0.26 & 0.9 & 0.14 &  35.634 & 2.78 & 0.20 & $5.57\cdot 10^{34}$& $5.69\cdot 10^{-05}$ \\
12 & J1857+0212 & 3.7 & 16 & 52 & 97.7  & -0.28 & 0.9 & 0.14 &  34.345 & 8.58 & 1.64 & $3.01\cdot 10^{32}$& $3.07\cdot 10^{-07}$ \\
13 & J1902+0556 & 3.2 &  9 & 42 & 82.9  & -1.26 & 0.8 & 0.11 &  33.088 & 3.93 & 9.18 & $7.93\cdot 10^{31}$& $6.82\cdot 10^{-08}$ \\
14 & J1915+1009 & 3.1 &  9 & 33 & 85.8  & -0.54 & 0.7 & 0.11 &  33.960 & 5.37 & 4.19 & $3.16\cdot 10^{32}$& $2.70\cdot 10^{-07}$ \\
15 & J1917+1353 & 3.0 & 15 & 24 & 80.3  &  1.85 & 0.7 & 0.24 &  34.586 & 4.07 & 4.28 & $2.33\cdot 10^{33}$& $1.88\cdot 10^{-06}$ \\
16 & J1926+1648 & 3.1 & 13 & 44 & 90.5  & -0.39 & 0.6 & 0.13 &  33.562 & 7.75 & 5.11 & $6.07\cdot 10^{31}$& $5.18\cdot 10^{-08}$ \\
17 & J1932+22   & 2.2 &  4 &  8 & 44.0  &  0.65 & 0.6 & 0.21 &  35.877 & 9.31 & 0.40 & $8.69\cdot 10^{33}$& $5.33\cdot 10^{-06}$ \\
18 & J2002+3217 & 3.5 &  4 & 25 & 78.7  & -1.49 & 0.6 & 0.09 &  34.087 & 6.55 & 1.05 & $2.85\cdot 10^{32}$& $2.75\cdot 10^{-07}$ \\
\hline
& $\Sigma_{1-18}$   & 51.0 & 288 & 896 & 1228.8 & -0.54  & - &0.036   &        &    &     &  $8.30\cdot 10^{-5}$ & \\
\hline
\end{tabular}
\label{tab_psr_selected}
\end{center}
\end{table*}
For 19 SNRs, with $l>20^\circ$ and with
an estimated distance $d<10$ kpc (Case and Bhattacharya, 1998)
statistics was accumulated (see Table \ref{tab_snr_selected}), 
resulting in an equivalent observation time of 52.3 h.
A significance of 1.1 standard deviations was derived and an
upper limit of 6.7\% of the Crab flux was calculated.\\
Theoretical predictions for the hadronic gamma-ray emissivity of
SNRs are taken from Drury, Aharonian \& V\"olk (1994), hereafter DAV:
\begin{equation}
F_{\gamma}(>E) \approx 9 \times 10^{-11}\Theta \left(\frac{E}{1
    TeV}\right)^{-1.1} \times S~ cm^{-2} s^{-1}
\label{eq1}
\end{equation}
The scaling value S is given by:
\begin{equation}
S=\left(\frac{E_{SN}}{10^{51}
    erg}\right)\left(\frac{d}{1 kpc}\right)^{-2}\left(\frac{n}{1 cm^{-3}}\right) 
\label{eq2}
\end{equation}

Assuming for the product of the fraction of mechanical energy converted to
cosmic rays times the SNR energy $\Theta E_{SN}$ a value of $1.6\cdot
10^{50}$ erg as given e.g. by Fields et al. (2001),
placing the remnant at a distance $d = 1$ kpc and assuming an average 
value of $n = 1$ atom per cm$^3$ for the interstellar medium density, 
a flux $F_{SNR}$ of 78\% of the
remnant compared to the flux of the Crab ($F_{Crab}$) is
calculated in the energy range between 1 and 10 TeV.
Following the paper of DAV lower flux ratios of 48\% and 18\% are
derived for softer SNR spectra of index $-1.2$ and $-1.3$. \\
We note, that the value for $\Theta E_{SN}$ is quite uncertain,
depending on the rate of SN explosions in our Galaxy (see e.g.
Dragicevich, Blair and Burman (1999)) on the mass of interstellar
matter in the galaxy and on the spectral index of the source
population of CRs.
Berezhko \& V\"olk (2000b) derive a value between $0.5 \cdot 10^{50}$ and $3.3
\cdot 10^{50}$ erg from CR energetics arguments. Fields et al. (2001) calculate
from similar arguments a value of $1.6 \cdot 10^{50}$ erg but cannot
exclude values as low as $1 \cdot 10^{50}$ erg and as high as $10
\cdot 10^{50}$ erg.
A similar value of $1 \cdot 10^{50}$ erg to $3 \cdot 10^{50}$ erg is derived in
Drury, Markiewicz \& V\"olk (1989). 
Theoretical calculations for diffusive shock accelerating
models 
based on spherical symmetry of the acceleration process
tend to give a value for $\Theta$ on the level of 50\% or even
higher (see e.g. Drury, Markiewicz \& V\"olk (1989) and consequently Drury, Aharonian \& 
V\"olk (1994)),
resulting in a value of $5\cdot10^{50}$ erg for a standard SNR of
$10^{51}$ erg.  
\\
An additional uncertainty arises from the density of the interstellar medium 
in which the SNR evolves. Berezhko \& V\"olk (2000a) point out that a 
large fraction of the Galaxy is occupied
by the so called hot interstellar medium where the particle density is
lower than 1 atom per cm$^3$, leading to low $\pi^0$-gamma-ray 
production in such an ambient medium. 
\\
For comparison of 
the derived ensemble limit with theoretical predictions, the difficulty of
estimating the distance to individual remnants complicates the
situation. 
Only for a minority of SNRs reliable estimates for the distance exist from
direct measurements (see e.g. Green 1998). A recent estimate
of the distance to most of the known remnants is given in Case and
Bhattacharya (1998) using the radio-surface-brightness-to-diameter 
relationship ($\Sigma$-$D$). In the calculations here we adopt their values. 
The limit derived by the observation can be compared with the 
predictions of the DAV model by summing
over all remnants given in Table \ref{tab_snr_selected} and by
weighting the individual
source by the fractional observation time $T_i$.
Thus we define the flux $F_{DAV}$ predicted by the DAV model as follows:
\begin{equation}
\frac{F_{DAV}}{F_{Crab}} = \sum_i \frac{F_{\gamma}^i}{F_{Crab}} \frac{T_i}{T}= \frac{F_{SNR}}{F_{Crab}} \cdot \frac{1 kpc^2}{T}\sum_i{\frac{T_i}{d_i^2}} 
\label{equ_ensemble_sum}
\end{equation}
\begin{table*}[t]
\begin{center}
\caption{Results for SNRs and pulsars
not used in the ensemble limits of
Table. \ref{tab_snr_selected} and Table \ref{tab_psr_selected}. Columns are labeled as in
Table. \ref{tab_snr_selected} and in the same units.}
\vspace{0.04cm}
\scriptsize
\begin{tabular}{|r|l|r|r|r|r|r|r|r||r|l|r|r|r|r|r|r|}
\hline
No & Name & $r$     & T   & ON & OFF & $\sigma$ & E$_{th}$ & $F^{99\%}$   &  No & Name & T   & ON & OFF & $\sigma$ & E$_{th}$ & $F^{99\%}$  \\
   &      & [$^\circ$] & [h] &    &     &          & [TeV]    & [CU]      &     &      & [h] &    &     &          & [TeV]    & [CU]        \\
\hline
\hline
\multicolumn{9}{|l||}{SNRs}                                   & 30& J1818-1422 &3.8 & 53 & 156 & 0.12 & 1.8 & 0.34 \\     
\cline{1-9}
 20& G359.0-0.9 & 0.2 & 6.6 & 1043 & 3028 & 0.9  & 4.8 & 9.90 & 31& J1820-1346 &2.1 & 38 & 115 & -0.05  & 1.8 & 0.54 \\  
 21& G359.1-0.5 & 0.3 & 7.1 & 1105 & 3449 & -1.1 & 4.7 & 4.68 & 32& J1822-1400 &2.1 & 34 & 124 & -1.03  & 1.8 & 0.40 \\  
 22& G359.1+0.9 & 0.1 & 5.1 & 821 & 2500 & -0.4  & 4.7 & 8.54 & 33& J1823-1115 &1.6 & 21 & 69 & -0.37   & 1.6 & 0.55 \\  
 23& G000.0+0.0 & 0.0 & 7.1 & 1156 & 3410 & 0.5  & 4.5 & 7.74 & 34& J1824-1118 &2.3 & 34 & 86 & 0.83    & 1.6 & 0.55 \\  
 24& G000.3+0.0 & 0.2 & 7.1 & 1118 & 3356 & -0.0 & 4.4 & 5.94 & 35& J1825-1446 &1.6 & 18 & 50 & 0.28    & 1.8 & 0.58 \\  
 25& G000.9+0.1 & 0.1 & 3.6 & 482 & 1738 & -3.6  & 4.2 & 2.30 & 36& J1826-1131 &2.6 & 36 & 107 & 0.05   & 1.6 & 0.41 \\  
 26& G001.0-0.1 & 0.1 & 4.6 & 726 & 2178 & 0.0   & 4.2 & 4.96 & 37& J1826-1334 &2.6 & 53 & 164 & -0.20  & 1.7 & 0.50 \\  
 27& G001.4-0.1 & 0.1 & 1.5 & 312 & 904 & 0.5    & 4.1 & 9.59 & 38& J1827-0958 &1.6 & 14 & 34 & 0.66    & 1.5 & 0.47 \\  
 28& G001.9+0.3 & 0.0 & 1.0 & 207 & 633 & -0.2   & 4.0 & 9.87 & 39& J1832-1021 &2.6 & 14 & 60 & -1.25   & 1.5 & 0.20 \\  
 29& G008.7-0.1 & 0.4 & 1.6 & 274 & 737 & 1.5    & 2.7 & 3.49 & 40& J1837-06   &3.3 & 35 & 79 & 1.37    & 1.3 & 0.34 \\  
 30& G009.8+0.6 & 0.2 & 2.1 & 80 & 340 & -2.9    & 2.5 & 0.52 & 41& J1836-0436  &1.8 & 6 & 47 & -2.50   & 1.2 & 0.20 \\  
 31& G010.0-0.3 & 0.1 & 2.1 & 65 & 220 & -0.9    & 2.5 & 0.67 & 42& J1842-03    &2.6 & 13 & 60 & -1.47  & 1.1 & 0.17 \\ 
 32& G011.2-0.3 & 0.0 & 2.1 & 63 & 212 & -0.8    & 2.4 & 0.68 & 43& J1844-0433  &2.6 & 25 & 63 & 0.73   & 1.1 & 0.34 \\ 
 33& G011.4-0.1 & 0.1 & 2.3 & 62 & 218 & -1.1    & 2.4 & 0.53 & 44& J1848-0123  &3.7 & 36 & 65 & 2.38   & 1.0 & 0.34 \\ 
 34& G012.0-0.1 & 0.1 & 1.8 & 62 & 205 & -0.7    & 2.3 & 0.79 & 45& J1852+00    &5.8 & 32 & 78 & 0.97   & 0.9 & 0.16 \\ 
 35& G013.5+0.2 & 0.0 & 1.3 & 58 & 167 & 0.3     & 2.1 & 1.35 & 46& J1857+0057  &3.2 & 10 & 47 & -1.35  & 0.9 & 0.11 \\ 
 36& G015.9+0.2 & 0.1 & 4.9 & 82 & 297 & -1.5    & 1.9 & 0.22 & 47& J1901+0156  &1.7 & 3  & 14 & -0.73  & 0.9 & 0.18 \\ 
 37& G016.7+0.1 & 0.0 & 2.1 & 30 & 105 & -0.8    & 1.8 & 0.41 & 48& J1901+0331  &2.6 & 11 & 47 & -1.09  & 0.8 & 0.14 \\ 
 38& G016.8-1.1 & 0.3 & 1.0 & 31 & 85 & 0.4      & 1.8 & 0.95 & 49& J1908+04    &2.1 & 4  & 23 & -1.29  & 0.8 & 0.14 \\ 
 39& G018.8+0.3 & 0.2 & 2.6 & 63 & 219 & -1.0    & 1.7 & 0.41 & 50& J1902+06    &3.2 & 12 & 36 & 0.00   & 0.8 & 0.16 \\ 
 40& G018.9-1.1 & 0.3 & 1.0 & 51 & 133 & 0.8     & 1.7 & 1.53 & 51& J1905+0709  &3.2 & 17 & 54 & -0.21  & 0.8 & 0.16 \\ 
 41& G020.0-0.2 & 0.1 & 2.1 & 36 & 100 & 0.4     & 1.6 & 0.56 & 52& J1906+0641  &3.2 & 17 & 42 &  0.67  & 0.8 & 0.18 \\ 
 42& G023.6+0.3 & 0.1 & 2.3 & 23 & 54 & 1.0      & 1.3 & 0.37 & 53& J1901+0716  &2.1 & 4 & 19 & -0.88   & 0.8 & 0.14 \\ 
 43& G024.7+0.6 & 0.3 & 2.8 & 54 & 194 & -1.2    & 1.3 & 0.27 & 54& J1902+07    &2.1 & 6 & 26 & -0.84   & 0.7 & 0.17 \\ 
 44& G027.4+0.0 & 0.0 & 2.6 & 27 & 75 & 0.3      & 1.2 & 0.33 & 55& J1908+0916  &3.4 & 12 & 44 & -0.63  & 0.7 & 0.11 \\ 
 45& G027.8+0.6 & 0.4 & 1.6 & 63 & 187 & 0.1     & 1.1 & 0.69 & 56& J1909+1102  &2.1 & 3 & 26 & -2.00   & 0.7 & 0.11 \\ 
 46& G029.7-0.3 & 0.0 & 3.1 & 21 & 60 & 0.2      & 1.1 & 0.21 & 57& J1908+07    &3.4 & 14 & 34 & 0.66   & 0.7 & 0.16 \\ 
 47& G031.5-0.6 & 0.2 & 2.6 & 28 & 110 & -1.3    & 1.0 &  0.19& 58& J1910+07    &2.6 & 7 & 25 & -0.41   & 0.7 & 0.14 \\ 
 48& G032.1-0.9 & 0.4 & 4.2 & 115 & 293 & 1.5    & 1.0 &  0.37& 59& J1912+10    &3.1 & 14 & 30 & 1.02   & 0.7 & 0.18 \\ 
 49& G033.2-0.6 & 0.2 & 5.7 & 54 & 167 & -0.2    & 1.0 & 0.14 & 60& J1913+09    &3.4 & 8 & 40 & -1.39   & 0.7 & 0.09 \\ 
 50& G036.6-0.7 & 0.3 & 1.6 & 43 & 100 & 1.4     & 0.9 & 0.59 & 61& J1914+1122  &2.6 & 15 & 34 & 0.89   & 0.7 & 0.23 \\ 
 51& G042.8+0.6 & 0.3 & 2.4 & 48 & 128 & 0.7     & 0.7 & 0.36 & 62& J1916+1030  &3.1 & 6 & 31 & -1.29   & 0.7 & 0.10 \\ 
 52& G054.1+0.3 & 0.0 & 2.6 & 10 & 22 & 0.8      & 0.6 & 0.20 & 63& J1916+0951  &2.1 & 6 & 16 & 0.24    & 0.7 & 0.17 \\ 
 53& G055.0+0.3 & 0.2 & 2.0 & 11 & 42 & -0.7     & 0.6 & 0.19 &	64& J1916+1312  &3.0 & 12 & 24 & 1.12   & 0.6 & 0.20 \\
 54& G057.2+0.8 & 0.2 & 2.5 & 11 & 27 & 0.6      & 0.6 & 0.22 &	65& J1918+1444  &3.0 & 8 & 20 & 0.43    & 0.6 & 0.16 \\
 55& G059.5+0.1 & 0.0 & 3.0 & 15 & 25 & 1.8      & 0.5 & 0.28 &	66& J1921+1419  &3.0 & 10 & 21 & 0.91   & 0.6 & 0.16 \\
 56& G059.8+1.2 & 0.2 & 3.0 & 11 & 36 & -0.2     & 0.5 & 0.17 &	67& J1918+15    &0.5 & 0 & 1 & -0.76    & 0.6 & 0.37 \\
 57& G063.7+1.1 & 0.1 & 1.5 & 2 & 8 & -0.4       & 0.5 & 0.21 &	68& J1926+1434  &3.0 & 8 & 28 & -0.39   & 0.6 & 0.13 \\
 58& G065.7+1.2 & 0.2 & 4.6 & 29 & 74 & 0.7      & 0.5 & 0.21 &	69& J1931+15    &2.1 & 7 & 13 & 0.99    & 0.6 & 0.24 \\
 59& G067.7+1.8 & 0.1 & 2.5 & 5 & 12 & 0.4       & 0.5 & 0.18 &	70& J1923+17    &2.1 & 8 & 18 & 0.66    & 0.6 & 0.21 \\
 60& G069.7+1.0 & 0.2 & 4.0 & 20 & 61 & -0.1     & 0.5 & 0.16 &	71& J1935+17    &1.6 & 2 & 10 & -0.70   & 0.6 & 0.16 \\
 61& G073.9+0.9 & 0.2 & 2.0 & 5 & 17 & -0.2      & 0.5 & 0.18 &	72& J1927+18    &2.6 & 1 & 15 & -1.98   & 0.6 & 0.09 \\
 62& G074.9+1.2 & 0.1 & 4.2 & 15 & 27 & 1.5      & 0.5 & 0.15 &	73& J1927+18    &2.6 & 7 & 19 & 0.22    & 0.6 & 0.17 \\
 63& G076.9+1.0 & 0.1 & 2.9 & 10 & 32 & -0.2     & 0.6 & 0.14 & 74& J1929+18    &2.6 & 7 & 21 & 0.00    & 0.6 & 0.16 \\
\cline {1-9}                                                       
\multicolumn{9}{|l||}{Pulsars}                                 & 75& J1926+19    &1.5 & 2 & 6 & 0.00    & 0.6 & 0.23 \\ 
\cline{1-9}                                                
 19& J1739-29   & &5.1 & 812 & 2474 & -0.38  & 4.8 & 8.61     & 76& J1932+2020  &2.8 & 5 & 21 & -0.70  & 0.6 & 0.14 \\    
 20& J1740-3015 & &4.1 & 681 & 2011 & 0.36   & 4.8 & 18.00    &	77& J1939+24    &2.0 & 2 & 8 & -0.37   & 0.5 & 0.16 \\  
 21& J1745-3040 & &4.6 & 698 & 1945 & 1.66   & 4.7 & 13.48    &	78& J1939+2134  &2.2 & 7 & 12 & 1.14   & 0.5 & 0.24 \\  
 22& J1749-3002 & &3.1 & 468 & 1328 & 1.03   & 4.7 & 9.13     &	79& J1946+2244  &1.0 & 4 & 4 & 1.52    & 0.5 & 0.52 \\  
 23& J1752-2806 & &1.5 & 271 & 922 & -1.84   & 4.0 & 5.43     &	80& J1946+26    &3.0 & 7 & 20 & 0.11   & 0.5 & 0.16 \\  
 24& J1803-2137 & &2.1 & 123 & 370 & -0.03   & 2.7 & 1.11     &	81& J1954+2923  &3.3 & 5 & 18 & -0.37  & 0.5 & 0.13 \\  
 25& J1808-20   & &1.6 & 67 & 226 & -0.85    & 2.6 & 0.91     &	82& J1955+2908  &4.6 & 5 & 30 & -1.55  & 0.5 & 0.07 \\  
 26& J1809-21   & &1.6 & 71 & 205 & 0.28     & 2.6 & 1.26     &	83& J2004+3137  &3.5 & 7 & 20 & 0.11   & 0.5 & 0.14 \\  
 27& J1812-1718 & &2.4 & 89 & 254 & 0.40     & 2.1 & 1.04     &	84& J1948+3540  &1.2 & 5 & 6 & 1.47    & 0.5 & 0.44 \\  
 28& J1812-1733 & &2.4 & 100 & 267 & 0.99  & 2.1 & 1.15       &	85& J2013+3845  &1.6 & 5 & 12 & 0.41   & 0.5 & 0.26 \\  
 29& J1816-17   & &1.6 & 60 & 155 & 0.97   & 2.1 & 1.38       &	86& J2029+3744  &1.6 & 2 & 4 & 0.46    & 0.6 & 0.18 \\  
\hline
\end{tabular}
\label{tab_snr2}
\end{center}
\end{table*}
where $F_{\gamma}^i$ for each source $i$ is given by Eq. \ref{eq1} and Eq. \ref{eq2}
in the energy range between 1 and 10 TeV.
For $\Theta E_{SN}$ we assume a value of $1.6 \times 10^{50}$ erg/s on average and for the
density of the interstellar medium a value of 1 atom per cm$^3$ on average.
This calculation gives a flux of 0.029 CU for a spectral index of -1.1 
of the complete ensemble which is about a factor
of two below the derived limit of 0.067 CU.
This result is within errors still consistent but obviously rules out
a strong enhanced emission of the 
SNR population compared to the prediction of DAV. Despite the large
uncertainties in the averaged gamma-ray luminosity of SNRs as
discussed before, this limit could also be viewed as a limit on
non hadronic production channels of gamma-rays in SNRs.
\\
We note that the derived
flux from the ensemble is quite stable with respect to the distance
estimate to individual remnants. An alternative way to 
derive an estimate on the distance for individual remnants 
by assuming a Sedov-Taylor expansion of the shell into the
interstellar medium, assuming an average age of the population 
of $2\cdot10^4$ years and using the values for the shell size as 
given in Shu (1992) gives within a few percent the same value
for the flux on the ensemble.
\subsection{Pulsars}
Compared to the SNR population the situation for pulsars is less
complicated for the following reasons:\\
\\
- pulsars are for all practical purposes point sources for HEGRA\\
- the distances to the pulsars can be estimated by the dispersion
measure (DM) of the pulsed radio emission and by modeling the
thermal electron distribution in the
Galaxy\\

For further analysis, pulsars with a characteristic age $\tau = 1/2~p/\dot{p}$
less than $10^6$ years, a rotation period of less than 1 s and a
modeled distance of less than 10 kpc were selected. 18 out of 86
pulsars fulfill these criteria. A list of these pulsars is given in
Table \ref{tab_psr_selected}. A comparison of the
estimated flux for the selected pulsars compared to the estimated flux
for the Crab-Pulsar out of the spin down luminosity and the distance 
show that even the most energetic pulsar in the 
selected sample is well below three orders of magnitude compared to
the Crab pulsar. Aharonian, Atoyan \& Kifune (1997) point out that the 
efficiency in converting spin down energy into high energetic
gamma-rays depends inversely on the magnetic field B in the surrounding 
synchrotron nebula. Due to the high magnetic field in the Crab Nebula
the expected efficiency in producing gamma-rays could be significantly 
higher (up to a factor of hundred) for other pulsars with a low
magnetic field compared to the 
Crab Nebula. Even under such an optimistic assumption a detection
seems to be quite unlikely.
The situation does not improve for the ensemble of the 18
pulsars. Under the assumption of the same efficiency in converting spin 
down energy into TeV gamma-rays in the individual pulsars as in the
Crab pulsar, 
a flux $\approx 400$ times lower as the calculated upper limit on the
ensemble of 3.6\% of the Crab flux 
is expected.

\section{Summary}

In a systematic search for point sources in the Galactic plane in the 
longitude range from -2$^o$ to 85$^o$ with the HEGRA IACT system 
no TeV gamma-ray emission was detected on a level above
4.5$\sigma$ in a total observation time of 115 h. 
Upper limits for 63 SNRs, 86 pulsars and nine unidentified GeV-sources 
on the level between 7\% of the
Crab flux and up to 18 Crab flux units were derived, depending on 
observation time and zenith angle. Summation over the most promising 
sources for TeV gamma-ray emission within each source class did not
yield an indication for emission from the SNR ensemble,
the PSR ensemble or the ensemble of GeV-sources. For the ensemble of 7 GeV
sources an upper limit of 5.7\% compared to the Crab flux was derived. For 
the ensemble of 18 pulsars selected by characteristic age and distance a
similar upper limit of 3.6\% was produced.
A theoretical estimate for these pulsars using the same conversion
efficiency from rotational energy to gamma-rays as in the Crab Nebula
gives a flux of approximately a factor 400 lower than the derived limit. \\
For an ensemble of 19 selected SNRs a limit of 6.7\% of the Crab flux
was derived. Comparing this limit with
a predicted hadronic gamma-ray flux of 2.9\% according to the DAV 
model and reasonable parameters rules out a strong enhancement 
of the emission of the SNR
population compared to the model predictions.\\
While no new TeV sources could be established in this survey, we 
nevertheless note that with systems of IACTs such a survey provides an
efficient method to probe extended regions of the sky with a dense
population of sources, such as the Galactic plane.
 
\section*{Acknowledgments}

The support of the HEGRA experiment by the German Ministry for Research
and Technology BMBF and by the Spanish Research Council
CICYT is acknowledged. We are grateful to the Instituto
de Astrof\'\i sica de Canarias for the use of the site and
for providing excellent working conditions. We gratefully
acknowledge the technical support staff of Heidelberg,
Kiel, Munich, and Yerevan. 

\newpage
\clearpage


\begin{thebibliography}{}
%
\bibitem[Aharonian et al.(2001)]{2001A&A...370..112A} Aharonian, F.~et al.\ 2001a, A\&A, 370, 112
\bibitem{diffuse} Aharonian, F.\ A., et al., 2001b, A\&A 375, 1008
\bibitem{tycho} Aharonian, F.\ A., et al., 2001c, A\&A 373, 292
\bibitem[Aharonian et al.(2000)]{2000ApJ...539..317A} Aharonian, F.~A.~et al.\ 2000, ApJ, 539, 317
\bibitem{501_spect1} Aharonian, F., et al., 1999a, A\&A 342, 69
\bibitem{501_spect2} Aharonian, F., et al., 1999b, A\&A 349, 11
\bibitem{galactic_tev_sources} Aharonian, F.\ A., 1999c, Astropart. Phys. 11, 225
\bibitem{pulsars} Aharonian, F.\ A., Atoyan A.\ M. \& Kifune T., 1997, MNRAS 291, 162
\bibitem{Berezhko} Berezhko, E.\ G., V\"olk, H.\ J., 2000a, Astropart. Phys. 14, 201  
\bibitem{Berezhko1} Berezhko, E.\ G., V\"olk, H.\ J., 2000b, ApJ, 540, 923
\bibitem{snr_whipple} Buckley et al., 1998, A\&A 329,639
\bibitem{hegra_trig} Bulian, N., et al., 1998, Astropart. Phys. 8, 223
\bibitem{case} Case, G.\ L. \& Bhattacharya, D., 1998, ApJ, 504, 761
\bibitem{Dragicevich} Dragicevich, P.M., Blair D.G. and Burman R.R., 1999, \\MNRAS 302, 693
\bibitem{hegra_perf} Daum, A., et al., 1997, Astropart. Phys. 8, 1
\bibitem{drury-snr} Drury, L.O., Aharonian, F.A. \& V\"olk, H.J., 1994, A\&A 287, 959
\bibitem{drury_old} Drury, L.O., Markiewicz, W.J. \& V\"olk, H.J., 1989, A\&A 225, 179
\bibitem{Fields} Fields, B. et al., 2001, A\&A, 370, 623
\bibitem{green_cat} Green, D.A., A Catalouge of Galactic Supernova Remnants,
Mullard Radio Astronomy Observatory, 1998
\bibitem{hartmann2} Hartmann, R.C., et al., 1999, ApJSS 123, 79
\bibitem[1983]{upl} Helene, O., 1983, NIM 212, 319
\bibitem{hegra_camera} Hermann, G., 1995,
   Proceedings of the Int. Workshop ``Towards a Major Atmospheric 
   Cherenkov Detector IV'', Padua, M. Cresti (Ed.), p. 396
\bibitem{psr17044} Kifune, T. et al., 1995, ApJ 438, L91
\bibitem{hegra_mc} Konopelko, A., et al., 1999a, Astropart. Phys., 10, 275
\bibitem{large_zenith} Konopelko, A., et al., 1999b, J. Phys. G, 25, 1989
\bibitem[Lamb \& Macomb(1997)]{1997ApJ...488..872L} Lamb, R.~C.~\& Macomb, D.~J.\, 1997, ApJ, 488, 872
\bibitem{neural} Lampeitl, H. \& Konopelko, A., 1999, Proc. of the
  26th ICRC, Salt Lake City, 4, 81
\bibitem{lima} Li, T., Ma, Y., 1983, ApJ 272, 317
\bibitem[CT1]{Mirzoyan} Mirzoyan, R.~et al., 1994, NIM 351, 513
\bibitem[Muraishi et al.(2000)]{2000A&A...354L..57M} Muraishi, H.~et al., 2000, A\&A, 354, L57
\bibitem{hegra_pointing} P\"uhlhofer, G., et al., 1997, Astropart. Phys. 8, 101
\bibitem{hegra_survey} P\"uhlhofer, G, et al. 1999, Proc. 26th ICRC, 4, 77
\bibitem[Romero, Benaglia, \& Torres(1999)]{1999A&A...348..868R}
  Romero, G.~E., Benaglia, P., \& Torres, D.~F.\ 1999, A\&A 348, 868
\bibitem{w28} Rowell, G.P. et al., 2000, A\&A 359, 337
\bibitem{schaefer} Sch\"afer, B.M. et al., 2001, NIM A465, 394
\bibitem{Shu1992} Shu, F. H., 1992, The Physics of Astrophysics,
  Volume II, University Science Books, 233
\bibitem{cos-b} Swanenburg, B.N. et al.,1981, ApJ 243, L69-L73
\bibitem[Tanimori et al.(1998)]{1998ApJ...497L..25T} Tanimori, T.~et al.\ 1998, ApJL, 497, L25
\bibitem{tayler} Taylor, J.H. et al., 1993, ApJSS 88, 529
\bibitem{snr_hegra} V\"olk, H. J. in Proc. Towards a Major Atmospheric
  Cherenkov Detectoe-V, O.C. de Jaeger, ed. (Kruger National Park,
  1997), 87
\bibitem[Weekes et al.(1989)]{1989ApJ...342..379W} Weekes, T.~C.~et al.\ 1989, ApJ, 342, 379
\bibitem[Yoshikoshi et al.(1997)]{1997ApJ...487L..65Y} Yoshikoshi, T.~et al.\ 1997, ApJ, 487, L65
%
\end{thebibliography}
\end{document}